\documentclass[final,5p,times,twocolumn,authoryear]{elsarticle}

\usepackage{amssymb}
\usepackage{amsmath,bm}
\usepackage{multirow}

\usepackage{color}

\newcommand{\degree}{\ensuremath{^\circ}}

\journal{JGR-Planets}

\begin{document}

\begin{frontmatter}


\title{Image Reconstruction Techniques in Neutron and Gamma-Ray Spectroscopy:  Improving Lunar Prospector Data}




\author[1]{Jack T. Wilson}
\author[1]{David J. Lawrence}
\author[1]{Patrick N. Peplowski}
\author[1]{Joshua T. S. Cahill}
\author[2]{Vincent R. Eke}
\author[2]{Richard J. Massey}
\author[3]{Lu\'{\i}s F. A. Teodoro}

\address[1]{The Johns Hopkins University Applied Physics Laboratory, 11100 Johns Hopkins Road, Laurel, MD 20723, USA}
\address[2]{Institute for Computational Cosmology, Department of Physics, Durham University, Science Laboratories, South Road, Durham DH1 3LE, UK}
\address[3]{BAER, Planetary Systems Branch, Space Sciences and Astrobiology Division, MS 245-3, NASA Ames Research Center, Moffett Field, CA 94035-1000, USA}

%
%
%
%



\begin{abstract}
We present improved resolution maps of the Lunar Prospector Neutron Spectrometer thermal, epithermal and fast neutron data and Gamma-Ray Spectrometer Th-line fluxes via global application of pixon image reconstruction techniques. With the use of mock data sets, we show that the pixon image reconstruction method compares favorably with other methods that have been used in planetary neutron and gamma-ray spectroscopy.  The improved thermal neutron maps are able to clearly distinguish variations in composition across the lunar surface, including within the lunar basins of Hertzsprung and Schr{\"o}dinger.  The improvement in resolution reveals a correlation between albedo and thermal neutron flux within the craters. The consequent increase in dynamic range confirms that Hertzsprung basin contains one of the most anorthositic parts of the lunar crust, including nearly pure anorthite over a region tens of km in diameter.  At Orientale, the improvement in spatial resolution of the epithermal neutron data show that there is a mismatch between measures of regolith maturity that sample the surface and those that probe the near-subsurface, which suggests a complex layering scenario. 
\end{abstract}

%
%
%

\end{frontmatter}



\section{Lunar Gamma-ray and Neutron Spectroscopy}
The global distribution of all major and several minor elements on the Moon was mapped by the Lunar Prospector (LP) Gamma-Ray Spectrometer (GRS) and Neutron Spectrometer (NS) \citep{Elphic98,Elphic2000,Feldman98,Feldman2000,Feldman2002,Lawrence1998,Lawrence2000,Lawrence2002,Prettyman2006}. Information in these maps has given rise to an improved
understanding of the formation and evolution of the lunar surface and
interior (e.g. \citealt{Jolliff2000,Hagerty2006,Hagerty2009}). Unlike other spectroscopic techniques, both gamma ray and neutron spectroscopy have the advantage of being able to provide bulk measurements of elemental composition at depths of a few 
tens of centimeters, which is orders of magnitude deeper than that available from ultraviolet to near-infrared spectroscopy.

For gamma rays detected from the
natural decay of thorium (Th), uranium and potassium, the
inferred abundances do not depend on cosmic ray flux or the abundance of other elements \citep{Metzger77}.  Because these ions preferentially partition into the melt phase during igneous processing, their abundance acts as an indicator of past 
magmatic activity and differentiation.

Planetary neutron data are typically separated into three energy ranges: high-energy, fast neutrons (500 keV $\lesssim$ E $\lesssim$ 10 MeV); intermediate energy, epithermal 
neutrons (0.3 eV $\lesssim$ E $\lesssim$ 500 keV) and slower, thermal neutrons (E $\lesssim$ 0.3 eV). Epithermal neutron data are sensitive to the presence of hydrogen and the LP-NS 
epithermal data have revealed the presence of subsurface hydrogen at the poles \citep{Feldman2000,Feldman2001}, which is preferentially concentrated within the cold traps in 
permanently shaded regions \citep{Feldman2001,Eke2009,Teodoro2010}.  Thermal neutron data provide a map of neutron absorbing elements, which on the Moon are predominately Fe, Ti and the rare 
earth elements (REE) Gd and Sm \citep{Elphic2000}.  Analysis of these data to create maps of absolute neutron absorption also revealed the presence of nearly pure plagioclase in parts of the 
highlands \citep{Peplowski2016}.  The fast neutron data provide a measure of mean atomic mass within the top metre of lunar regolith 
\citep{Maurice2000,Gasnault2001}, because the higher neutron to proton ratio in Fe and Ti leads to a higher fast neutron flux.

In addition to their ability to measure bulk elemental abundance, the penetrating power of gamma rays and neutrons leads to the footprint of the point spread functions (PSFs) of orbital GRS and neutron 
NS to be similar to the spacecraft's altitude, thus spatial resolution is improved by lowering orbital altitude.  This relatively large PSF hinders comparison
with other, better-resolution, remotely-sensed data, such as visible imaging, due to the averaging effects of observation with the GRS/NS.  This averaging also results 
in a reduction of the dynamic range of the measured data.  

An efficient method of reducing these complications with the GRS/NS data is the application of image reconstruction techniques 
to sharpen the maps and suppress the effects of noise.  Several image reconstruction or deconvolution techniques have been used with GRS/NS data from several missions, including Jansson's method on Mars Odyssey Neutron Spectrometer (MONS) epithermal data at Tharsis \citep{Elphic2005} and the Martian poles to monitor CO$_{\rm 2}$ cap growth and retreat \citep{Prettyman2009}; the pixon method on LP Th data \citep{Lawrence2007,Wilson2015}, LPNS epithermal data to map hydrogen within permanently shadowed craters at the lunar poles \citep{Elphic2007,Eke2009} and the global MONS epithermal data to map near-subsurface water on Mars \citep{Wilson2018}; and a new method, called here, the Regularized Maximum Likelihood method to examine the near-subsurface H distribution from MONS epithermal data \citep{Pathare2018}.

In this paper we apply image reconstruction methods to the global, lunar neutron and gamma-ray data to improve the utility of the mapped data. Additionally several 
deconvolution and image reconstruction methods, commonly used in planetary sciences, are compared to find that most suitable to the problem.  In the following section we 
offer a brief discussion of image reconstruction in planetary neutron and gamma-ray spectroscopy. More detailed explanations can be found in the references provided. 
Section~\ref{sec:valid} describes tests, carried out using mock data, of the various image reconstruction and deconvolution methods. Section~\ref{sec:data} details the data used in this study. The results of applying image reconstruction methods to these LP data are presented in Section~\ref{ssec:pixonRes}. We conclude in Section~\ref{sec:conc}.

\section{Image reconstruction in planetary sciences}\label{ssec:imageRec}
Observation by an imperfect instrument can often be modelled by stating that the data obtained, $D$, are the result of the convolution of the true, underlying, image, $I$, that 
would be measured with a perfect instrument, with the PSF of the actual detector, $B$, and the addition of noise, $N$, i.e.
\begin{equation}\label{eqn:obs}
D({\bf x}) = I*B({\bf x}) + N({\bf x}),
\end{equation}
where the argument ${\bf x}$ refers to position or pixel number (as appropriate) and $*$ is the spatial convolution operator. Solving for the inverse of equation~(\ref{eqn:obs}) to 
find the reconstruction closest to the original image, $I$, is the problem of image reconstruction.  As the noise is known only statistically, equation~(\ref{eqn:obs}) cannot be 
inverted analytically, so we must instead adopt some statistical or iterative technique to perform the inversion. 

Given that the number of degrees of freedom that may be involved in the problem is as large as the number of pixels that make up the image, searching significant regions of the 
parameter space can be computationally expensive.  Several techniques have been developed to combat this problem including direct inversion using iterative procedures such as 
Jansson's method \citep{Jansson1984}.  However, these techniques often lead to the production of spurious features in the reconstruction, driven by, and amplifying, noise in the data 
\citep{Lawrence2007}.  To avoid these problems Bayesian image reconstruction techniques have been developed that make use of prior information, favouring simplicity, in addition to 
the measured data.  The Bayesian technique most often employed in the reconstruction of planetary neutron and gamma ray data is the pixon method \citep{PP93,Eke2001}.  Below, we provide a 
brief description of some commonly used reconstruction techniques, that will be examined in this paper.

\subsection{Smoothing}
The most straightforward method of regularizing the data is to smooth the maps via convolution with a kernel.  For simplicity this kernel is usually chosen to be the PSF 
\citep{Lawrence2003}, however any function may be selected.  This technique does not attempt to remove the effect of blurring due to observation, but only to suppress 
statistical noise. Ideally, the kernel is chosen such that it is large enough to average away noise variations but no so large as to wash out real variations in the detected signal.

\subsection{Jansson's method}
If one considers the problem of deconvolution (neglecting the presence of noise) the most immediate solution is to find the inverse of the PSF and convolve this with the data. However, the PSF will often 
be band-limited so its inverse will be infinitely large.  This fact along with the presence of noise can give rise to reconstructions that are dominated by large, spurious and often 
oscillatory features. A series of linear, iterative algorithms have been developed with the aim of controlling this undesired behaviour as it evolves.

Jansson's method is one of these algorithms. It contains an additional relaxation factor, $r$, modifying each iteration, which is intended to suppress 
spurious noise-derived features that arise in linear deconvolution techniques. The method is defined by the relations   
\begin{align}
	\hat{I}^{(0)}({\bf x}) &= D*B({\bf x}), \\
	\hat{I}^{(k+1)}({\bf x}) &= {I}^{(k)}({\bf x}) + r[I^{(k)}](D({\bf x}) - {I}^{(k)}*B({\bf x})), \\
	r[I^{(k)}({\bf x})] &= r_0\left(1 - 2\left|\frac{I^{(k)}({\bf x})- I_{\rm min}}{I_{\rm max} - I_{\rm min}} - \frac{1}{2}\right|\right),
\end{align}
where $r_0$ is a relaxation constant whose value is determined empirically and $I_{\rm min}$ and $I_{\rm max}$ are the smallest and largest permitted values of the reconstruction. That is, the new guess at the truth, $\hat{I}^{(k+1)}$, is the sum of the old guess, $\hat{I}^{(k)}$, and the residuals (corresponding to the $k$th guess) scaled by a relaxation factor, $r$. The initial guess, $\hat{I}^{(0)}$, is set equal to the data convolved with the instrumental PSF.   The form of $r$ prevents solutions that are moving out of the allowed range of values from changing further, as it goes to zero in pixels where the solution approaches $I_{\rm min}$ or $I_{\rm max}$. It was hoped that avoiding non-physical solutions (e.g. negative fluxes) would improve the final image, even in regions outside of those truncated \citep{Jansson1984}. This method has seen some use in planetary sciences for the spatial deconvolution of neutron and gamma-ray data sets \citep{Elphic2005,Elphic2005b,Lawrence2007,Prettyman2009}. However it is still liable to produce spurious, noise-derived, features in the deconvolved images \citep{Lawrence2007}.

As repeated iterations of Jansson's method amplify noise, an appropriate maximum number of iterations must be found for each data set.  In practice this is done by creating a mock set of data with statistics and, ideally, features that closely match the actual data. The method is then applied to this mock data and the number of iterations required to find the reconstruction most closely matching the known truth is determined.  The method is then run on the actual data for the number of iterations found to be the most appropriate for the mock data.

\subsection{Regularized Maximum Likelihood}
A recent scheme for attempting to perform deconvolution was introduced in \citet{Pathare2018}.  This method involves describing the reconstructed image in spherical harmonic space and varying the spherical harmonic coefficients to minimize the measured reduced chi-squared, $\chi^2_r$, value based on the difference between the reconstruction, blurred with the instrumental PSF, and the data.  In this scheme the effects of noise on the reconstruction are suppressed by truncating the spherical harmonic power series to a given degree.  The maximum degree used is that necessary to reduce $\chi^2_r$ to $\sim$1.

\subsection{Pixon reconstruction}
Unlike direct inversion methods, whose aim is deconvolution with the suppression of noise a secondary consideration, Bayesian image reconstruction techniques start by asking a different question: given the data and our prior knowledge what is the most likely true, underlying image? 

Answering this question requires maximizing the posterior
conditional probability:
\begin{equation}\label{eqn:prob}
P(\hat{I},M|D) = \frac{P(D|\hat{I},M)P(\hat{I}|M)P(M)}{P(D)},
\end{equation}
where $\hat{I}$ is the inferred truth and $M$ the model
describing both how the image is represented in pixel space and our knowledge of equation~\ref{eqn:obs}. For our purposes $P(D)$ is simply a normalization factor and $P(M)$ will be taken as uniform. Thus two terms are of interest; the first, $P(D|\hat{I},M)$, is the likelihood of
the data given a particular inferred truth and model, which can be calculated using a goodness-of-fit statistic. For data with Gaussian errors $\chi^2$ can be used, where
\begin{equation}\label{eqn:like}
P(D|\hat{I},M) = \exp(-0.5\chi^2),
\end{equation}
and 
\begin{equation}
\chi^2 = \sum\left(\frac{D({\bf x}) - I*B({\bf x})}{\sigma({\bf x})}\right)^2,
\end{equation}
where $\sigma({\bf x})$ is the expected noise in pixel ${\bf x}$. The second term of interest, $P(\hat{I}|M)$, is the image prior, i.e. the probability of obtaining a particular inferred truth given the model. 

Many Bayesian image reconstruction techniques assume that the model, $M$, in equation~(\ref{eqn:prob}) is fixed.  However, this is not a necessary assumption: Allowing both $I$ and $M$ to vary in the reconstruction enables the prior to be interpreted as a mathematical statement of Occam's razor.  More complex models, which allow a larger number of reconstructed truths, are naturally disfavoured and allowed only if they provide a sufficiently good improvement in the likelihood.  In the pixon method, the model, i.e. the pixel arrangement, is allowed to vary. 

It was noted by \citet{PP93} that each cell, or pixel, represents a degree of freedom and in regions of the image with low signal to noise ratio or containing little structure 
this may cause the image to be over-specified.  Grouping sets of pixels together would remove this problem and greatly improve the prior, which is strongly 
dependent on the number of cells.  It is these groups of pixels, called pixons, that form the fundamental image units in pixon reconstruction.
For an image consisting of $n$ pixons and containing $C$ separate and indistinguishable
detections, the probability 
of observing $I_i$ detections in each pixon $i$ is
\begin{equation}\label{eqn:prior}
P(\hat{I}|M) = \frac{C!}{n^C \Pi^n_{i = 1}I_i!}.
\end{equation}

The problem of image reconstruction is then reduced to finding the pixon basis that restricts the prior while also allowing an acceptable likelihood to be found. In the Maximum 
Entropy Pixon (MEP) method the prior is maximized perfectly, for a given number of pixons, by having the same
information content in each pixon, i.e. $I_i = C/n$ for all $i$ \citep{PP93}. The most probable reconstruction is then found by varying the number of pixons to maximize the posterior probability 
(equation~\ref{eqn:prob}).  Alternatively the Locally Adaptive Pixon (LAP) method uses the definition of a local misfit statistic, ${E_R^\prime}$ (\ref{ssec:LocalE_R}), to 
determine how the pixon sizes must be modified at each location \citep{Wilson2018}.  

In both pixon methods the maximization of the posterior proceeds in a two-stage iteration.  First, for a given pixon distribution, the best reconstruction is found by minimizing the 
global misfit statistic, ${E_R}$ (\ref{ssec:E_R}). ${E_R}$ is chosen over the more commonly used $\chi^2$ as it penalizes correlations in the residuals, which helps to prevent the 
emergence of spurious features in the reconstruction.  Second, the pixon distribution is altered.  In the MEP scheme this involves adjusting the number of pixons in the 
reconstruction.  In the LAP case the pixon size distribution is modified directly, based on ${E_R^\prime}$.  This modification provides more freedom so that the pixon size is changed only where required, unlike in the MEP case where it must be varied globally to fit any local feature. The combination of global 
misfit statistic minimization and pixon size modification is repeated until the posterior probability is maximized.

The pixon method \citep{PP93} has successfully been used
in a range of disciplines including medical imaging and infrared  and
X-ray astronomy (\citealt{P96} and references therein). In addition, it
has recently been used to reconstruct remotely sensed neutron
\citep{Eke2009,Wilson2018} and gamma ray data \citep{Lawrence2007,Wilson2015} and has been
shown to give a spatial resolution $1.5$-$2$ times that of other
methods in reconstructing planetary data sets \citep{Lawrence2007}.  A modified pixon technique in which the reconstructed image is the result of combining decoupled regions has been developed for use when sharp boundaries are expected to exist based on other data \citep{Eke2009,Wilson2015}.  However, as we are concerned here with general reconstructions, not relying on external data, we will not explore this extension here.

\subsection{Wiener deconvolution}
The Wiener filter was designed to reduce noise.  Wiener deconvolution is an extension of this technique to the problem of deblurring \citep{Bracewell58,Helstrom67}.  Although the original motivation was to minimize the mean-square error in the reconstruction, Wiener deconvolution turns out to be the optimal Bayesian reconstruction when both the signal and noise are Gaussian random variables.  

In practice the filtering is performed in the spectral domain where the filter is given by \citep{Zaroubi95}
\begin{equation}
\tilde{W} = \frac{\tilde{B}^\star C_I}{|\tilde{B}|^2C_I + C_N},
\end{equation}
where the tilde indicates the spectral domain (i.e. Fourier or spherical harmonic transform), $^\star$ denotes complex conjugation, and $C_I$ and $C_N$ are the power spectra of the signal and noise, respectively.

As Weiner deconvolution consists of a single convolution, it is the fastest of the deconvolution methods discussed here. However, for the Wiener deconvolution technique to be effective, the covariance, or power spectra, of the signal and noise must be known before the deconvolution is performed, which will often not be the case.

\section{Validation of image reconstruction techniques using mock data}\label{sec:valid}
\subsection{Mock Planetary Data Sets}\label{ssec:mockData}
To illustrate the effect of image reconstruction on noisy data sets, we will create a series of mock data sets that simulate observations.  The true image in these mock data are based on Lunar Orbiter Laser Altimeter (LOLA) 1064 nm albedo data \citep{Lucey2014}. These data have complete coverage and are not affected by latitude-based effects, as are imaging data relying on solar illumination. Two classes of mocks are created: one with complete coverage and Gaussian noise with a set signal-noise-ratio (SNR), and a second with coverage and noise-amplitude equivalent to that of the LP-NS fast and GRS Th data. Nine mock data sets were created with complete coverage, with SNR of 100, 10 and 5, each with three different sized PSFs selected to approximate those of LP (full width at half maximum, FWHM, $\sim 45$\,km), Mars Odyssey Neutron Spectrometer (FWHM $\sim 550$\,km) and Dawn's Gamma Ray and Neutron Spectrometer. 
The PSF, $B^\prime$, was chosen to be a kappa function as described in \citet{Lawrence2003}. That is 

\begin{align}\label{eqn:B}
	B^\prime({\bf x};h) &= A\left(1 + \frac{{\bf x}^2}{2\sigma(h)^2}\right)^{-\kappa(h)-1},\\
	\sigma(h) &= 0.704h + 1.39,\\
	\kappa(h) &= -4.87\times10^{-4}h + 0.631,
\end{align} 
where $h$ is the detector altitude. Some of these data are shown in the top panels in Figures~\ref{fig:pixTest30}
~and~\ref{fig:pixTestTrue}.


\subsection{Validation using mock data}\label{ssec:mockResults}

\begin{figure*}
\begin{center}
\includegraphics[width=1.0\textwidth]{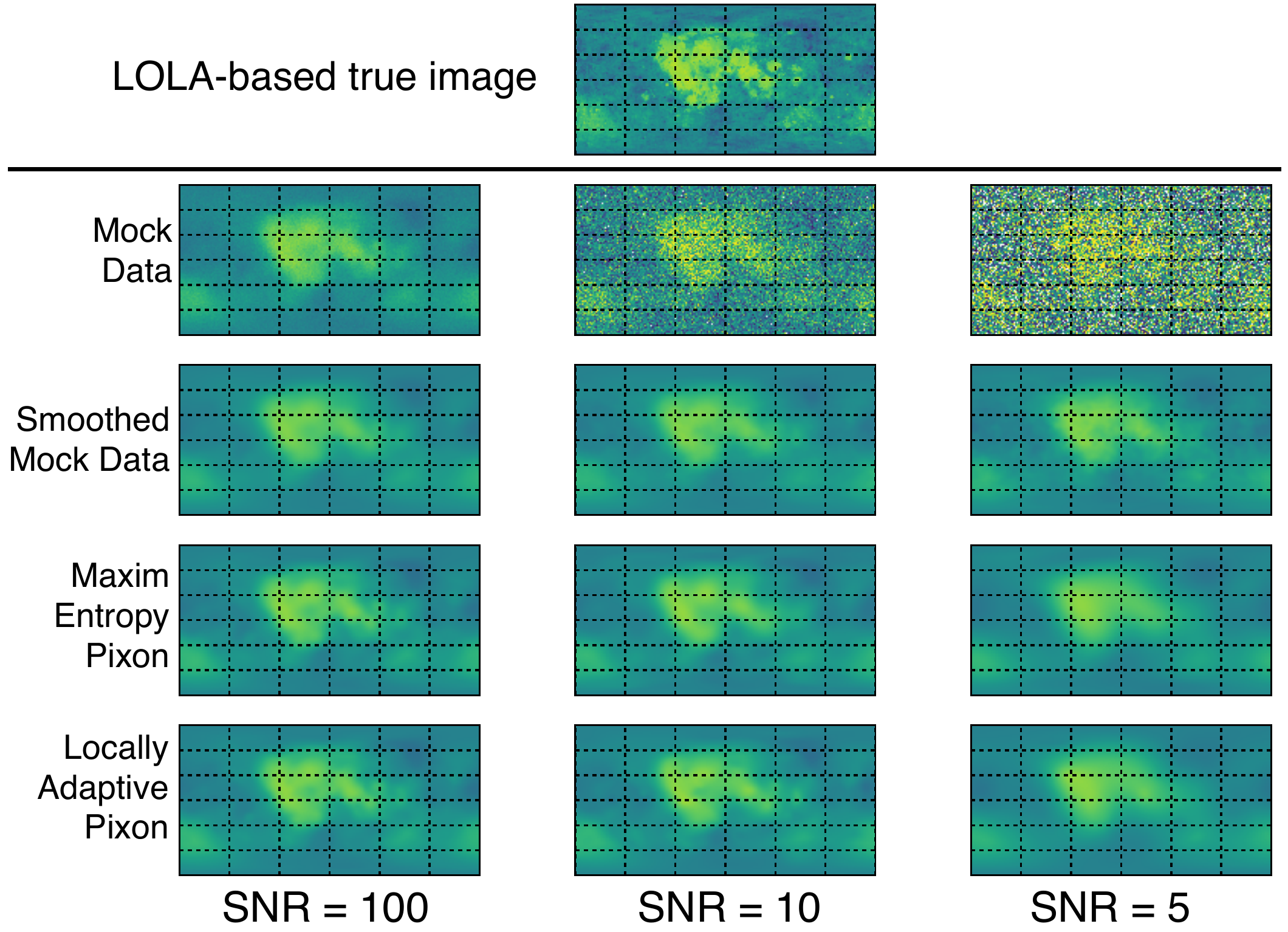}
\end{center}
\caption{Global maps of the true image, mock data sets and reconstructed versions of the data.  The top image shows the true image on which the mock data are based, and is a scaled map of the LOLA 1024 nm reflectance data \citep{Lucey2014}.  The other rows are in three columns each based around a mock data set with different signal to noise ratios.  The SNR decreases going left to right from 100 to 10 and 5.  The rows from the second going down show the mock data, PSF-smoothed mock data, Maximum Entropy Pixon, and Locally Adaptive Pixon reconstructions of the mock data sets.}
\label{fig:pixTest30}
\end{figure*}

\begin{figure}
\begin{center}
\includegraphics[width=1.0\columnwidth]{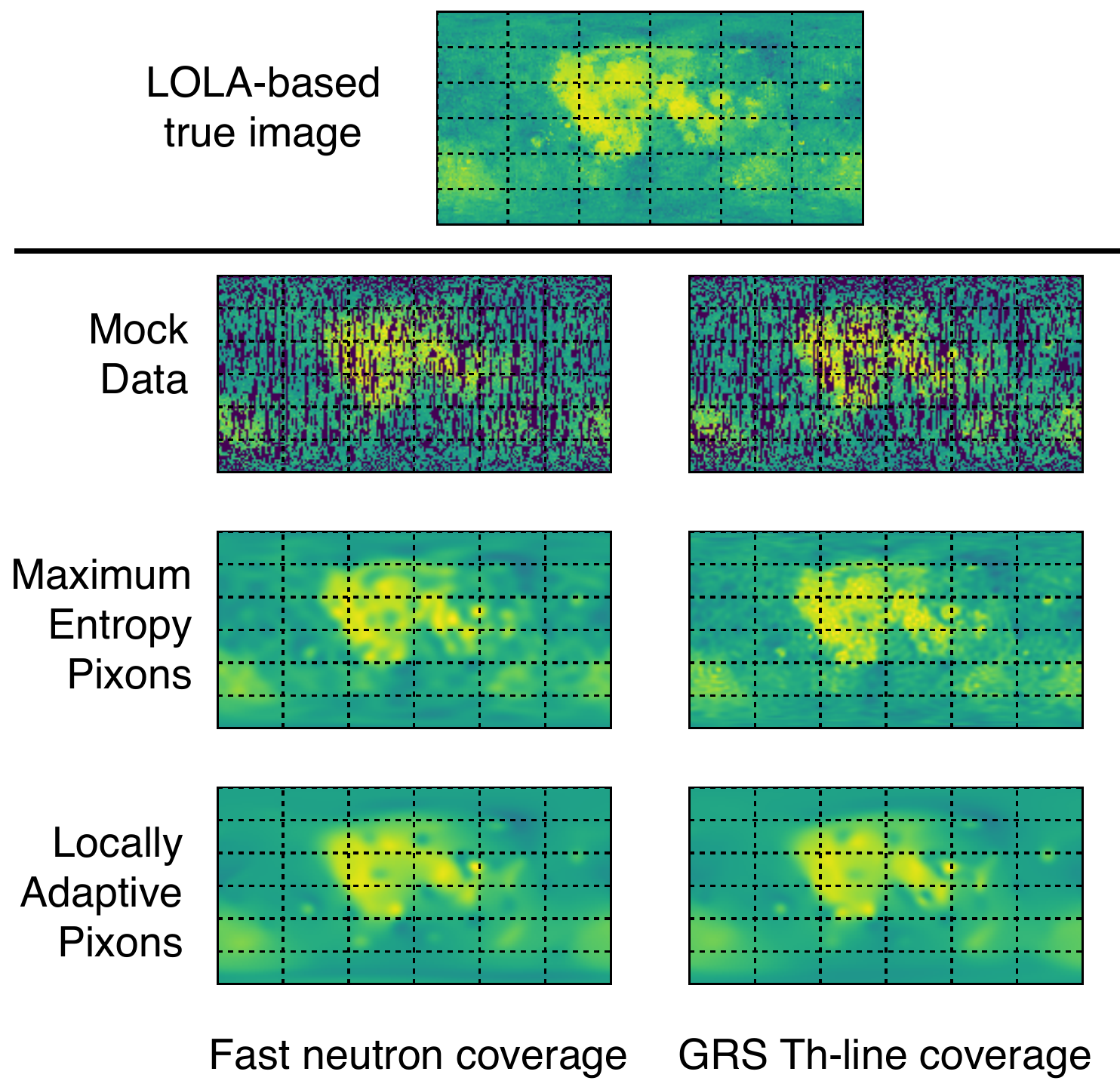}
\end{center}
\caption{Global maps of the true image, mock data sets and reconstructed versions of the data.  The top panel shows the true image on which the mock data are based, and is a scaled map of LOLA 1024 nm reflectance data \citep{Lucey2014}.  The other rows are in two columns each based around a mock data set with different coverage. The left and right columns have the same coverage as the fast neutron and GRS Th-line data, respectively.  The rows from the second going down show the mock data, Maximum Entropy Pixon, and Locally Adaptive Pixon reconstructions of the mock data sets.}
\label{fig:pixTestTrue}
\end{figure}

\begin{figure}
\begin{center}
\includegraphics[width=1.0\columnwidth]{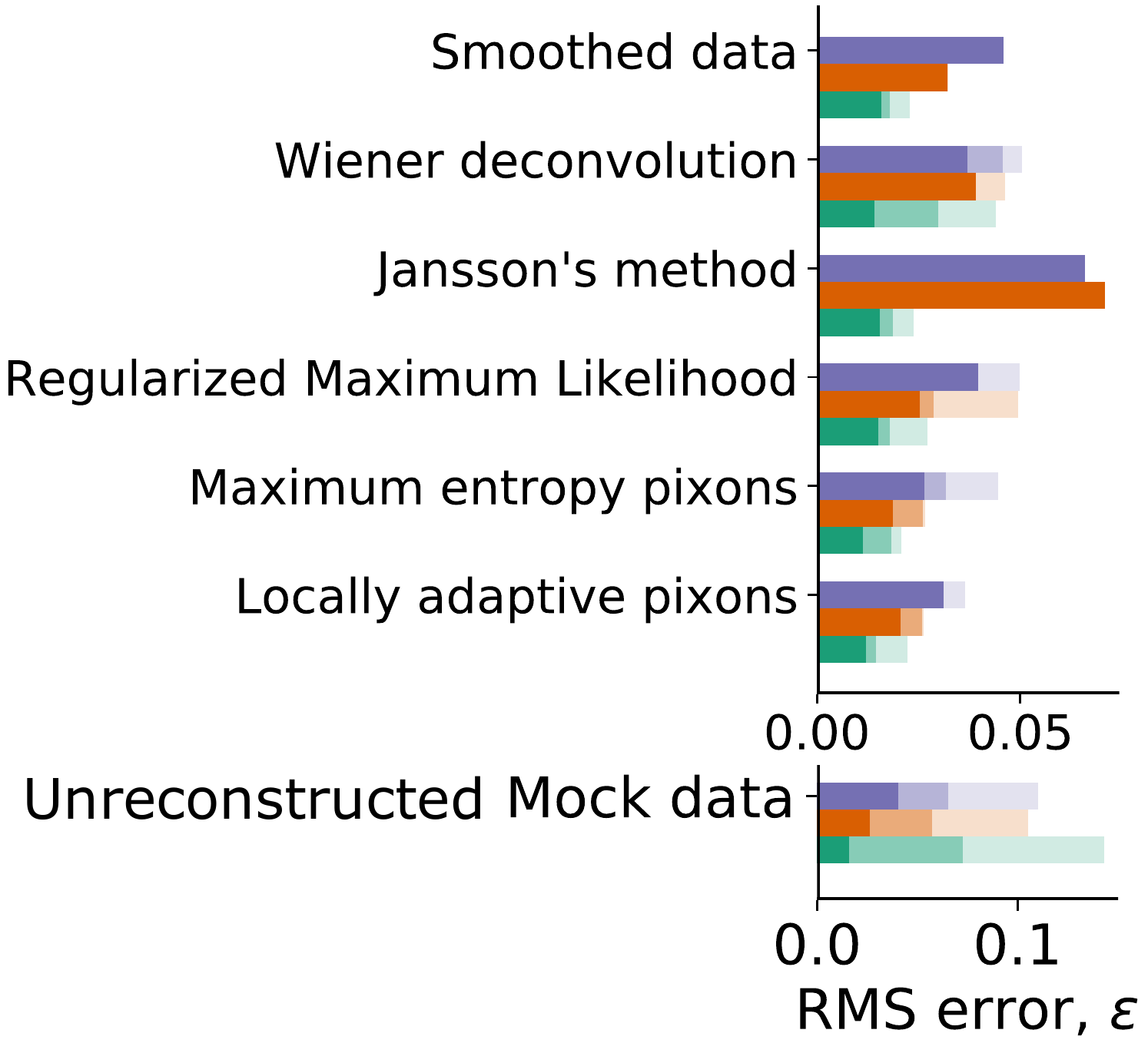}
\end{center}
\caption{The RMS difference between various reconstruction methods and the true image, i.e $\epsilon$ defined in equation~(\ref{eqn:eps}). Low values indicate accurate image reconstruction. The reconstructions are labelled and described in section~\ref{ssec:imageRec}.  The shades of the same colour correspond to different SNR, with 100 being most solid and 5 most transparent.  The colours indicate different PSFs with the bottom, green being LP-like; middle, orange bar MONS-like and top, purple bar Dawn-like.}
\label{fig:bars}
\end{figure}

The mock data sets described in section~\ref{ssec:mockData} were reconstructed using the methods described in section~\ref{ssec:imageRec}.  Some of these results are illustrated in Figures~\ref{fig:pixTest30}~and~\ref{fig:pixTestTrue}.
  The achieved reduction in noise for all three methods is immediately obvious.  The improvment in resolution is more easily observed when considering small features in profile (Figure~\ref{fig:smallCrossSec}). On global scales, which are large compared with the PSF, the effect of the reconstruction techniques on the spatial resolution and dynamic range of the signal are not as apparent.
  Two broad criteria can be used to assess the effectiveness of each image reconstruction method, namely the reconstruction's accuracy and spatial resolution improvement.  These two criteria are described below.

\subsubsection{Quantifying Improvement in Accuracy}
One measure of the acuracy of each reconstructed image is the RMS difference between the reconstructions and the true image, i.e.
\begin{equation}\label{eqn:eps}
	\epsilon = \sqrt{\sum_{\bf x}\left(I({\bf x}) - \hat{I}({\bf x})\right)^2},
\end{equation}
where the sum is over all pixels.  This value is plotted for each of the image reconstruction techniques and mock data sets in Figure~\ref{fig:bars}.  Examination of this plot shows that Wiener deconvolution and Jansson's method, although offering some improvement for the smallest PSF case, compare poorly to a simple smoothing of the data. The Regularized Maximum Likelihood reconstruction shows a considerable improvement in the accuracy of the data but the pixon methods provide the closest reconstruction to the true image. 

\begin{figure}
\begin{center}
\includegraphics[width=1.0\columnwidth]{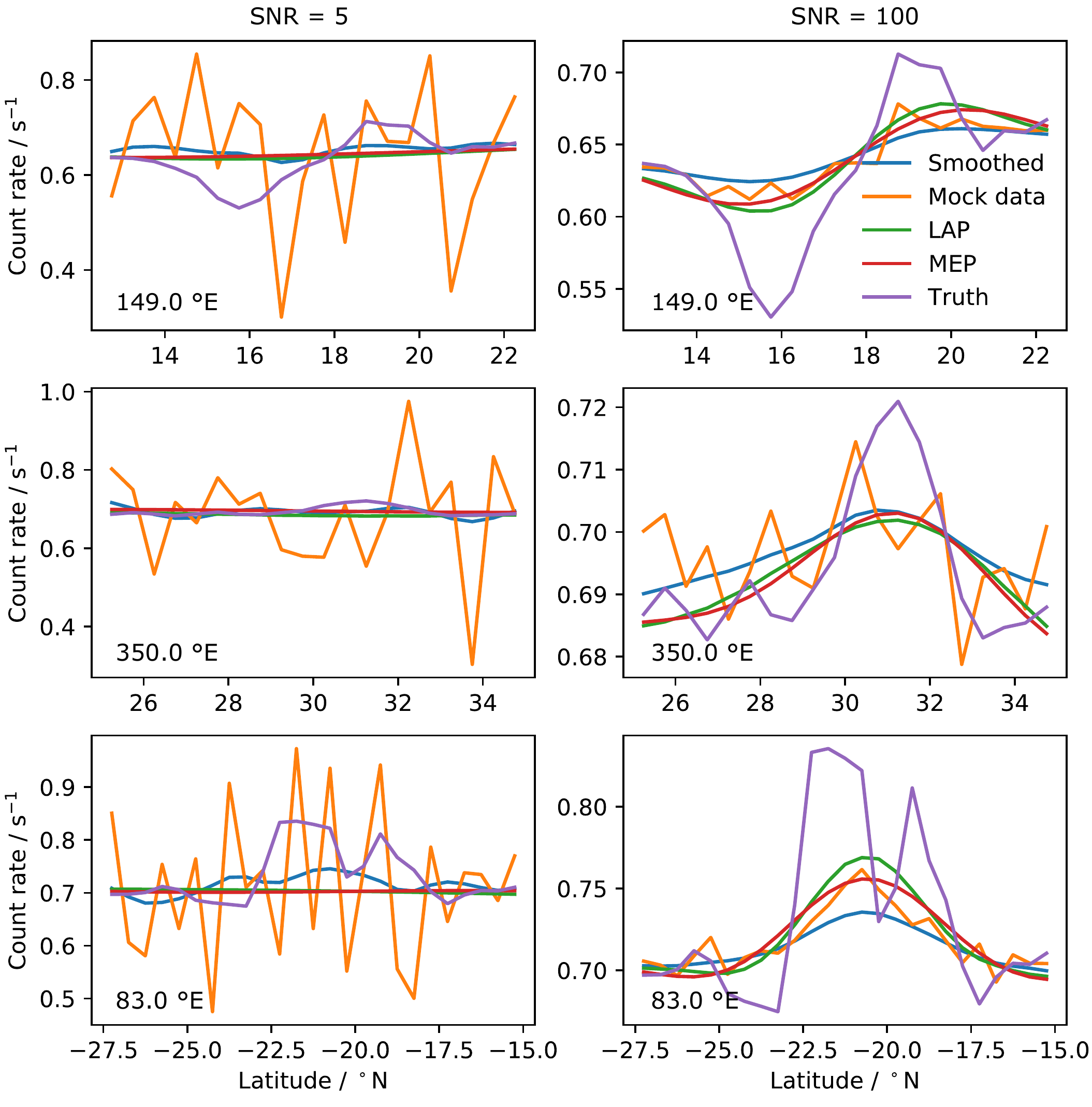}
\end{center}
\caption{Cross sections across the various data sets shown in Figure~\ref{fig:pixTest30}.  The cross sections all follow lines of constant longitude (indicated in each plot).  The two columns show data sets with different SNR (5 on the left and 100 on the right) and each row corresponds to a different location on the surface.}
\label{fig:smallCrossSec}
\end{figure}

Cross-sections through several small features, shown in Figure~\ref{fig:smallCrossSec} also provide information about reconstruction accuracy.  The different curves in Figure~\ref{fig:smallCrossSec} correspond to the following: 1) the true image chosen as the basis for the mock data; 2) the mock data themselves, which are the true image convolved with a PSF appropriate for observation from a 30 km orbit with added noise; 3) a smoothed map, in which the mock data have been smoothed with the PSF in order to suppress the effects of noise; and 4) MEP and LAP reconstructions of the mock data set.  These cross-sections provide a means to assess the relative performance of reconstruction techniques by comparing the reconstructions with the true image and smoothed data, which is the form in which neutron and gamma ray data are usually presented. 

Figure~\ref{fig:smallCrossSec} shows these data sets for two SNR runs and focuses on three locations containing features $\sim$ 60 - 150 km across.  The cross-sections shown in the figure are one pixel in width, which corresponds to 11 km at the equator.  For the high signal to noise ratio case (right hand column in Figure~\ref{fig:smallCrossSec}) it can be seen that, although, simply smoothing the data tends to preserve the features that are present and effectively average out the noise, the dynamic range is suppressed so the count rate in positive anomaly features is underestimated.  The reconstructed data sets also effectively suppress the noise but, typically, have the advantage that they increase the dynamic range of the counts, as compared with the raw data, and become closer to the true, underlying image than the smoothed data.  

The low signal to noise ratio data in the left hand column of Figure~\ref{fig:smallCrossSec} show another advantage of pixon reconstruction methods;  the avoidance of spurious features in the reconstruction.  In all of the low SNR panels, the pixon reconstructions show almost no variation in count rate over the cross-sections.  This is because the SNR is not sufficient to require the inclusion of the small features examined in the reconstruction.  In the case of the smoothed data some features still appear. However, there are also features that do not correspond to real variations in the true image but are simply artifacts introduced by the noise.  Thus, smoothed maps (and those derived by deconvolution or image reconstruction methods that are not adept at dealing with noise-derived artifacts) based on a real data set can contain features that are not physical.  However, the pixon methods avoid this shortcoming and all features that are present in the maps are really present on the surface.

Although the features present in the pixon reconstructions are required by the data and correspond to real variations in the underlying image, it is clear that pixon reconstruction techniques produce maps that have an intrinsic spatial resolution that varies across the image and is not equal to the original image.  This results in an effective smoothing of the truth and consequent loss in the definition of small scale features in the truth, including sharp boundaries.  Thus, the shapes of small features in the reconstructed map will most likely be smoother than is actually the case.  This should be kept in mind when attempting to match features in the reconstructed neutron and gamma ray data with other, higher-resolution data sets.

\subsubsection{Quantifying resolution}
\begin{figure*}
\begin{center}
\includegraphics[width=1.0\textwidth]{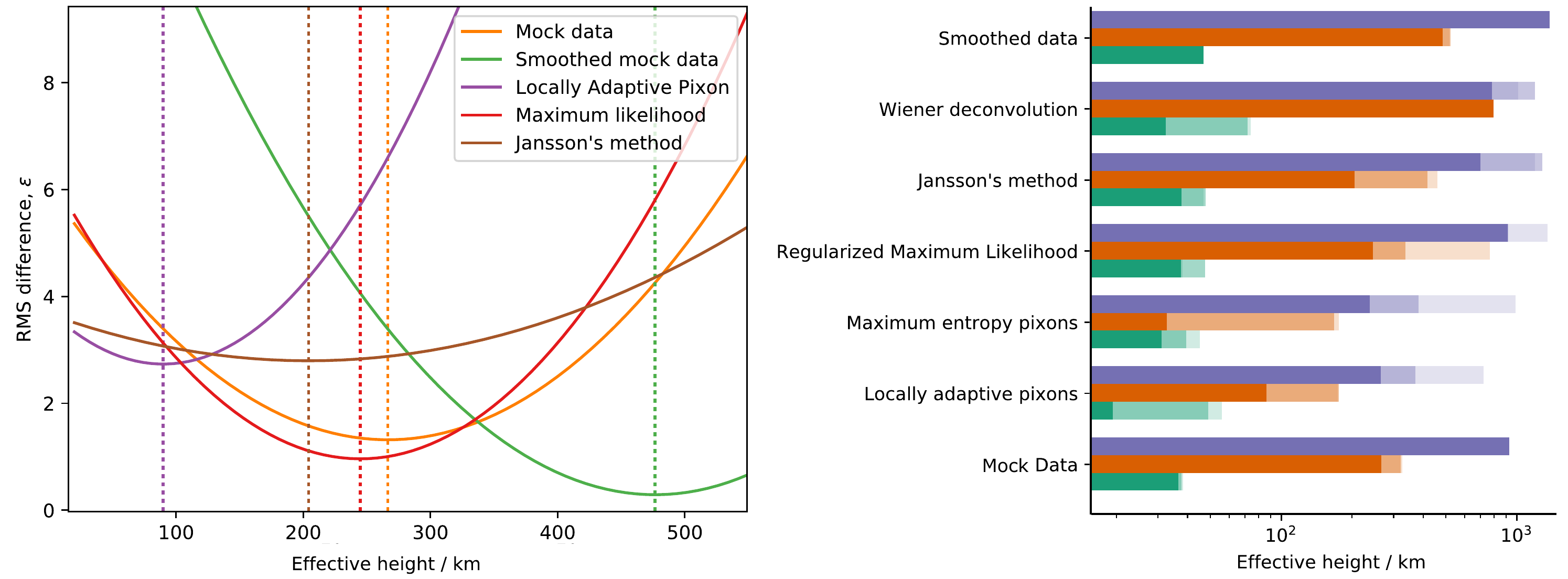}
\end{center}
\caption{{\it Left}. The misfit between the reconstruction and true image blurred by a PSF parameterized as an effective height.  The data used in this panel are the MONS-like mocks with SNR of 100.  The different colored curves correspond to different reconstruction methods, labelled in the legend.  The dashed lines indicate the locations of the minimum of each curve, which is equivalent to the effective resolution achieved by each technique. {\it Right}. Bar chart showing the resolution obtained in each mock reconstruction as determined by the method illustrated in the left panel. The shades of the same color correspond to SNR, with 100 being most solid and 5 most transparent.  The colours indicate different PSFs with the bottom, green being LP-like; middle, orange bar MONS-like and top, purple bar Dawn-like.}
\label{fig:res}
\end{figure*}

Previously, linear spatial resolution has been quantified by comparing the data and reconstruction power spectra \citep{Wilson2018}. However, when dealing with reconstructions of mock data sets their resolutions can be estimated more directly by determining by how much the original image must be blurred for it to most closely match the reconstructions.  This was calculated using the $\chi^2$-like statistic defined as
\begin{equation}\label{eqn:eps2}
	\epsilon^\prime = \sqrt{\sum_{\bf x}\left(I*B^\prime({\bf x})- \hat{I}({\bf x})\right)^2},
\end{equation}
where $B^\prime$ is the blurring function the size of which is varied to minimize $\epsilon^\prime$. For $B^\prime$ we chose to use the kappa function PSF, which is described in equation~\ref{eqn:B}. This allows for the translation of the resolution gain into an effective height to which the LP would have to descend to obtain an equivalent resolution.  The statistic, based on reconstructions of the LP-like (i.e. FWHM $\sim$45\,km) mock data using each technique described in section~\ref{ssec:imageRec} is shown in Figure~\ref{fig:res}. As expected smoothing the data increases the effective height of the detector (i.e. worsens the resolution), however its reduction of $\epsilon^\prime$ to close to zero shows that the blurred truth and smoothed data match well (i.e. smoothing effectively removes noise).  The Regularized Maximum Likelihood and Jansson's method have little effect on the resolution, however the absolute value of $\epsilon^\prime$ suggests that the Regularized Maximum Likelihood method is much more effective than Jansson's method at suppressing noise. Both pixon techniques provide a substantial improvement in resolution, in the MEP case the resolution is improved such that the effective height of the spacecraft is nearly halved from 36\,km to 19.5\,km. 

\subsection{Mock data summary}
Given the results of the previous two sections we conclude that the pixon methods provide the most effective tool for image reconstruction of remotely-sensed planetary data in terms of accuracy and resolution improvement.  It is these techniques that we will use to reconstruct the LP data sets.

\section{Pixon reconstruction of Lunar Prospector datasets}\label{sec:data}
In this study we will improve the utility of the LP neutron and gamma-ray (Th) data sets by applying image reconstruction techniques.  The LP data are first described, followed by the LP reconstructions and specific regional studies that illustrate different aspects of the reconstruction method.
\subsection{Overview of Lunar Prospector data}
Time-series LP-NS and GRS observations from the $7$ months the
Lunar Prospector spent in its low-altitude orbit (at $\sim 30$\,km) are used in this
work. 

The reduction of the neutron data is described in \citet{Maurice2004} and the gamma-ray data
in \citet{Lawrence2004}, with the counts in the Th decay line at $2.61$\,MeV 
being defined as the excess over a background value within the
$2.5-2.7$\,MeV range. Absolute Th abundances are determined following the
procedures used by \citet{Lawrence2003} and based in part on spectral unmixing work described in \citet{Prettyman2006}.

There are three NS derived data products. Epithermal neutrons are measured by a $^{\rm 3}$He proportional counter coated in a 0.63 mm Cd sheet to screen out thermal neutrons 
\citep{Feldman2004}. Thermal neutrons are measured as the difference between this $^{\rm 3}$He proportional counter and one coated in Sn, which does not screen thermal neutrons. 
Fast neutron measurements are provided by pulse-pair events in the GRS anti-coincidence shield \citep{Feldman2004}.

The PSFs of the two instruments were determined from functions described in \citet{Lawrence2003} and \citet{Maurice2004}, using the method in \citet{Wilson2015} to account for the motion of the spacecraft.

\subsection{Lunar Prospector Reconstruction results}\label{ssec:pixonRes}

\begin{figure*}
\begin{center}
\includegraphics[width=0.6\textwidth]{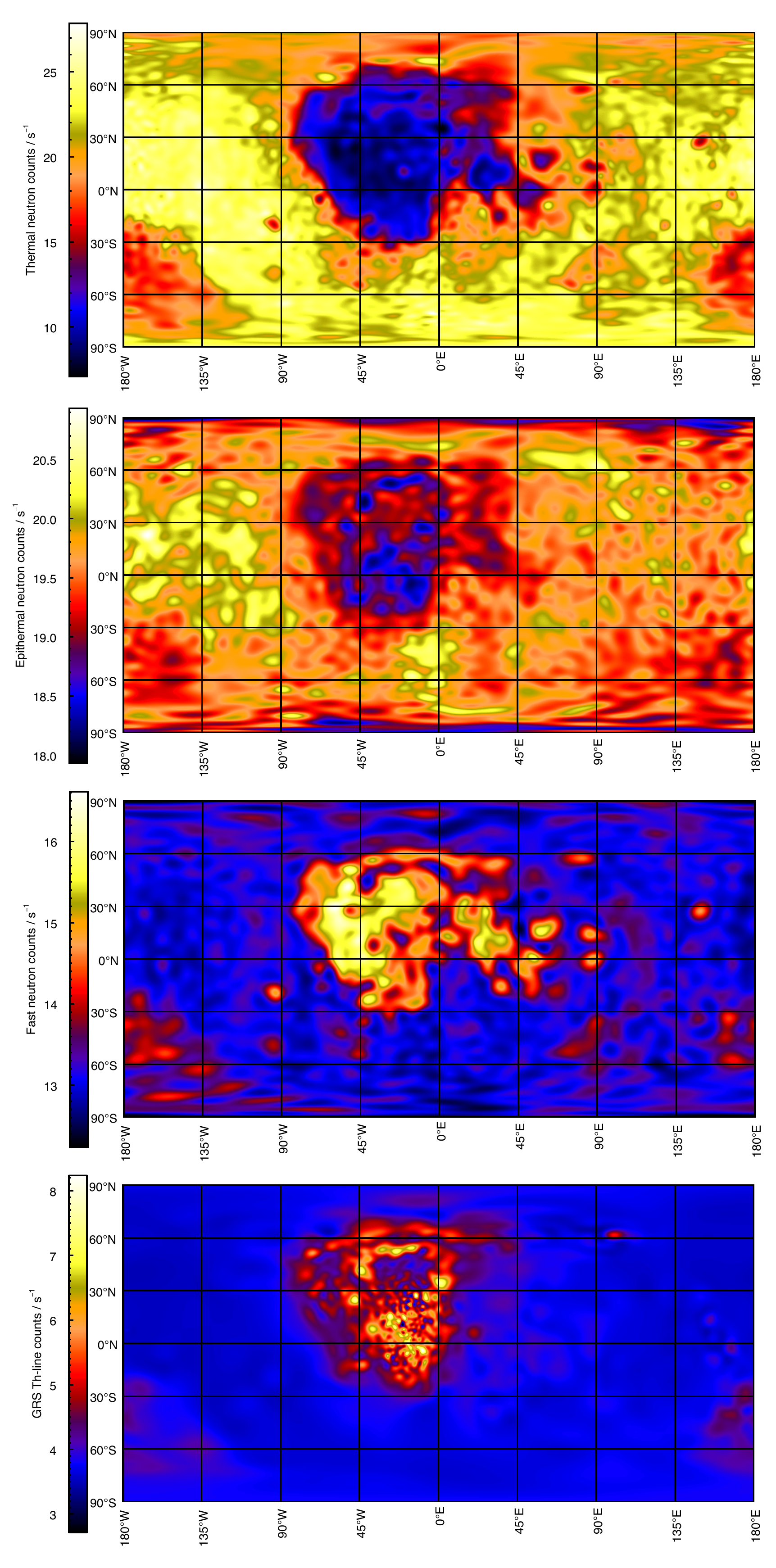}
\end{center}
\caption{{\it Top to bottom}:  Maximum Entropy Pixon reconstruction of the Lunar Prospector thermal, epithermal and fast neutron data and Locally Adaptive Pixon reconstruction of the Lunar Prospector GRS Th-line data.}
\label{fig:globals}
\end{figure*}

The results of the global pixon reconstructions of each of the LP data sets are shown in Figure~\ref{fig:globals}.  Maximum Entropy Pixon reconstructions of the neutron data are shown as is a Locally Adaptive Pixon reconstruction of the LPGRS Th-line data.  These reconstructions were chosen to illustrate both pixon methods. Small regions in each of these reconstructions are discussed below, including multi-ring basins of various sizes.  The examples below demonstrate, with a real data set, some of the effects explored in the previous section's mock reconstructions.

\subsection{Resolution improvement achieved}
\begin{figure}
\begin{center}
\includegraphics[width=\columnwidth]{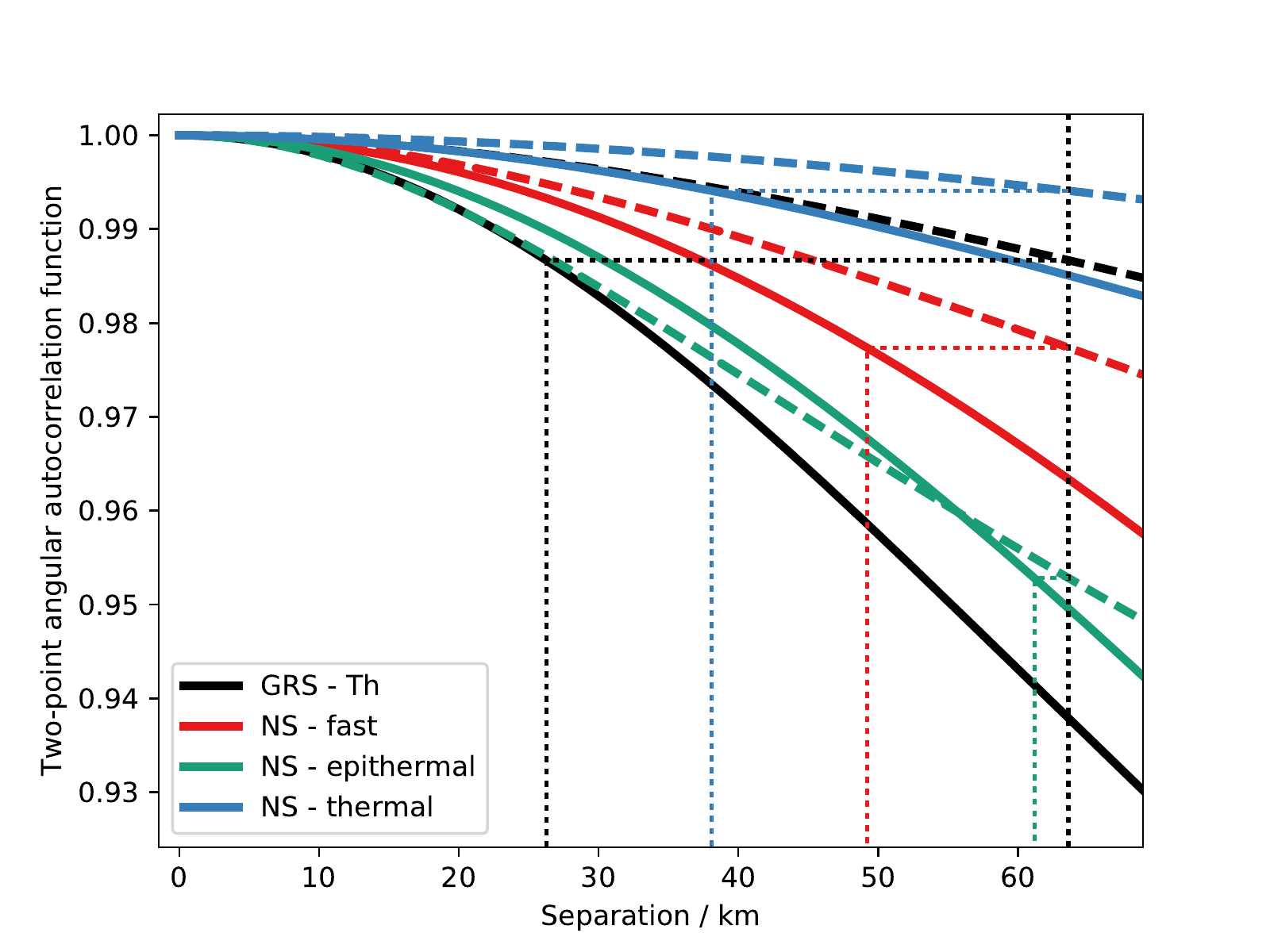}
\end{center}
\caption{The angular two-point correlation functions of the smoothed (dashed) and reconstructed (solid) Th-line GRS and thermal, epithermal and fast neutron fluxes.  The dotted lines indicate the separation at which the correlation functions of the reconstructions have fallen to the level that those of the PSF-smoothed data fall to at a separation equal to their resolution. Consequently, this provides a measure of the effective resolution of the reconstructions.}
\label{fig:2pt}
\end{figure}
Unlike the reconstructions in the previous section, based on mock datasets, the resolution of the reconstruction of the LP data cannot be estimated by direct comparison with the true image, which is unknown.  Instead, to estimate the resolution of the reconstructions, we compare the angular two-point autocorrelation functions of the data and reconstructions following \citet{Wilson2018}.  Due to the relatively low SNR of the LP-NS/GRS data compared with the Mars Odyssey Neutron Spectrometer data sets it was found to be necessary to first smooth the unreconstructed data with the instrumental PSF.  Without the PSF-smoothing the unreconstructed data show stong variation on small scales due to the noise in each pixel.  This would imply a much smaller resolution than is the case.

Figure~\ref{fig:2pt} shows that a significant improvement is seen in the resolution of the LP-GRS Th-line data, with the reconstruction having a resolution of 26\,km compared with the PSF-smoothed data's 63\,km resolution (determined by adding the FWHM of the PSF in quadrature).  The fast and thermal reconstructions show only small changes in resolution having two-point correlation function-derived resolutions of 49\,km and 38\,km respectively.  The epithermal reconstruction is seen to result in an increase in resolution to 61\,km.  This is consistent with the earlier results of \citet{McClanahan2010} and with pixon reconstruction providing an optimal smoothing when the SNR of the data does not allow for an improvement in spatial resolution.
%

\subsection{Investigation of specific regions}
To show the utility of these reconstructed maps, this section provides descriptions of five different regions that illustrate different aspects of spatial reconstruction methods given in Section~\ref{sec:valid}.  In some cases, resolution improvement is achieved that reveals new information about particular features.  In other cases, mapped data are made more accurate by minimizing spurious features.  A full understanding of these maps will require multiple detailed investigations that are beyond the scope of this paper.

\subsubsection{Hertzsprung Basin}\label{sec:Hertzsprung}

\begin{figure*}
\begin{center}
\includegraphics[width=\textwidth]{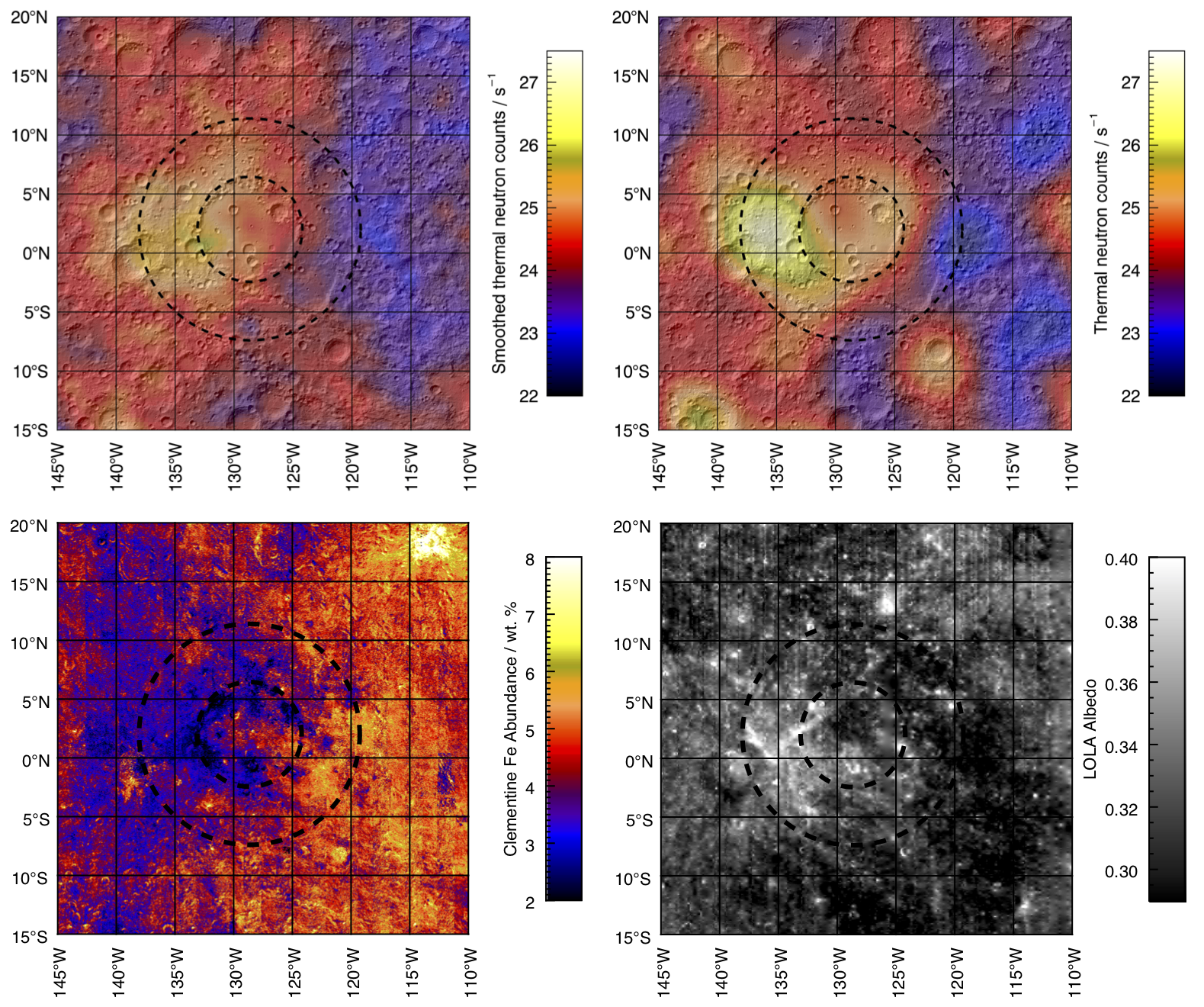}
\end{center}
\caption{{\it Clockwise from top left}: PSF-smoothed thermal neutron count rate at Hertzsprung. The Maximum Entropy Pixon reconstruction of the thermal neutron count rate at Hertzsprung. A LOLA 1064 nm albedo map of Hertzsprung.  Clementine-based FeO concentrations at Hertzsprung \citep{Lucey2000}. In all panels the black dashed lines indicate the inner and outer crater rims, with diameters of 270 km and 570 km. The neutron data are superimposed on a LOLA shaded relied image.}
\label{fig:Hertzsprung}
\end{figure*}

Hertzsprung is a well-preserved impact basin on the Moon's farside at 2\degree N, 128\degree W. With an outer rim diameter of 570 km it is an intermediate-sized lunar basin, transitional between two-ring basins, such as Schr{\"o}dinger (320 km diameter), and true multi-ring structures (i.e. those that clearly display 3 or more rings), such as Orientale \citep{Wilhelms1987}. Thus, it is expected to excavate material from depths intermediate between the two classes of features.  It has been estimated, from the crater's morphology, that the impact excavated material from as deep as 40 km below the surface, but with most ejecta sourced from shallower than about 25 km \citep{Spudis93}. The crust at Hertzsprung is estimated to be around 80 km thick \citep{Zuber94}, implying that material excavated by the Hertzsprung impactor samples the top half of the crust.  The crater contains two clear rings, the main rim of 570 km diameter and an inner ring of 270 km diameter (illustrated in Figure~\ref{fig:Hertzsprung}). 

Previous work examining Ultraviolet/Visible camera data from NASA's Clementine spacecraft has shown that the ejecta from Hertzsprung is unusually feldspathic \citep{Spudis96} and that the inner ring massifs contain almost completely iron free regions \citep{Stockstill98}, as seen in the lower left panel of Figure~\ref{fig:Hertzsprung}.  These Fe free regions were interpreted as outcrops of nearly pure anorthosite.  Hertzsprung is one of several basins, including Humorum, Nectaris and Orientale, that have exposed pure anorthosite massifs. These observations are consistent with pure anorthosite being a significant component of the crust at depths of several km, despite being uncommon on the surface. 

\citet{Peplowski2016} derived a map of absolute neutron absorption in the lunar highlands based on the high-altitude LP thermal neutron data. Within the Feldspathic Highlands Terrane (FHT) they identified several large regions whose low neutron absorption cross section implied the presence of high ($>85 wt.\%$) plagioclase abundance. These include regions co-located with the large basins Orientale, Hertzsprung, Birkoff, Korolev, Dirichlet-Jackson, Freundlich-Sharonov, and Mendeleev.

Hertzsprung has amongst the lowest neutron absorption on the surface and therefore also the highest abundance of plagioclase on the surface of the Moon.  It was observed by \citet{Peplowski2016} that the higher thermal neutron flux area extended beyond the crater on the western side, but the main area of low thermal neutron absorption was assumed to be coincident with the inner peak ring where Fe concentration is lowest (lower left panel Figure~\ref{fig:Hertzsprung}).  However the reconstructed thermal neutron data shows that the deficit and therefore the anorthosite abundant region is located in the inter-ring region (upper right panel Figure~\ref{fig:Hertzsprung}).  This region overlaps with the high albedo region identified in Lunar Reconnaissance Orbiter/Lunar Orbiter Laser Altimeter (LRO/LOLA) 1064 nm laser reflectance (lower right panel Figure~\ref{fig:Hertzsprung}). Compared with the unreconstructed data the thermal neutron count rate is higher in the feldspathic region, implying an even more Fe poor composition.  Thus the thermal neutron data reveal a large ($\sim$100 km) region containing nearly pure anorthosite. The basin is significantly mantled by deposits of the younger, Orientale basin, particularly in its southeastern sector.  This may be the reason for the decreased albedo and thermal neutron flux seen in the eastern half of the basin. 


\subsubsection{Schr{\"o}dinger Basin}
\begin{figure*}
\begin{center}
\includegraphics[width=1.0\textwidth]{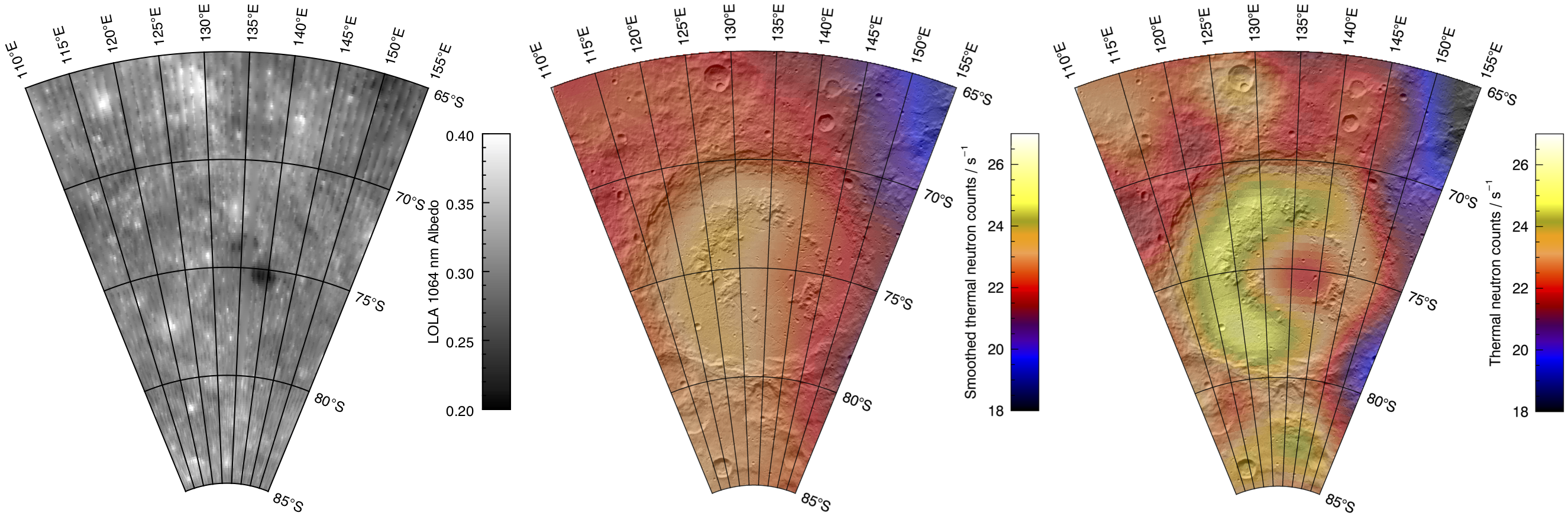}
\end{center}
\caption{{\it Left}: A LOLA normal albedo at 1064 nm map at Schr{\"o}dinger crater. {\it Center}: An unreconstructed, PSF-smoothed LPNS thermal neutron map. {\it Right}: A Maximum Entropy Pixon reconstruction of the Lunar Prospector thermal neutron data at Schr{\"o}dinger crater.  The neutron data are superimposed on a LOLA shaded relied image.}
\label{fig:Schrodinger}
\end{figure*}

The Schr{\"o}dinger basin, located at 75\degree S, 135\degree E, is somewhat smaller than Hertzsprung with a diameter of $\sim$ 320 km, but it too contains a peak ring, with a diameter of $\sim$ 150\,km (Figure~\ref{fig:Schrodinger}).  Like Hertzsprung this peak ring has been confirmed as containing km-scale regions of pure or nearly pure anorthosite on the basis of spectral data \citep{Kramer2013}.  Unlike Hertzsprung, Schr{\"o}dinger contains a large pyroclastic deposit \citep{Mest2011}, visible as a darker region in the left panel of Figure~\ref{fig:Schrodinger} at 75\degree\,S, 140\degree\,E, with the vent from which the deposit was sourced observable in the shaded relief in the right hand panel.  The effect of performing a pixon reconstruction on the data is seen by comparing the panels in Figure~\ref{fig:Schrodinger}.  Before reconstruction the separate features in the basin cannot be resolved, after reconstruction the pyroclastic deposit and western limb of the peak ring are clearly seen.  The absence of a clear eastern limb in the thermal neutron data is either the result of its proximity to the pyroclastic deposit and the resolution limitations of even the reconstructed data or an indication that the western limb is more plagioclase rich.  Indeed, there are a greater number of spectral detections of pure anorthosite in the west of the peak rim \citep{Kramer2013}. However, at such low latitudes the angle of solar incidence places constraints on the relative error of spectral data.


\subsubsection{Orientale Basin}

Orientale is located in the lunar highlands on the southwestern limb of the lunar nearside.  At $\sim$930 km diameter, it is the youngest large impact basin on the Moon. Due to its young age it is considered the standard example of a well preserved, multiring basin in comparative studies of large impacts on terrestrial planetary bodies. The basin exhibits at least three concentric topographic rings (including the Inner Rook, the Outer Rook, and the Cordillera rings) in a concentric arrangement around the Inner Depression. The Inner Depression is a central topographic low associated with the zone of excavated crust that extends to  160 km from the basin center and is bounded by a scarp.

\citet{Lawrence2015} showed that there is a general correlation in the lunar highlands between various measures of regolith maturity and epithermal neutron-derived H abundances.  The maturity measures considered were Clementine-derived 750 nm albedo, the highlands variation of which is attributed to long-term modifications of lunar soils by solar wind and micrometeorite bombardment \citep{Lucey2000}; optical maturity (OMAT), defined as an empirical relationship of the 750 nm and 950 nm reflectance ratios \citep{Lucey2000}; and LRO Diviner-derived wavelength position of the Christiansen Frequency \citep{Greenhagen2010}, which provides diagnostic information about bulk silicate polymerization and is found to be correlated with maturity \citep{Greenhagen2010,Allen2012}.  Epithermal neutrons are strongly moderated by implanted hydrogen, so an increase in epithermal flux corresponds to a decrease in implanted hydrogen abundance.  One interpretation of the correlation between these measures of maturity and epithermal neutron-derived H is that lunar surfical H abundance increases with exposure time due to solar wind protons being embedded into the regolith \citep{Lawrence2015}, however further study would be required to confirm this. 

\begin{figure*}
\begin{center}
\includegraphics[width=\textwidth]{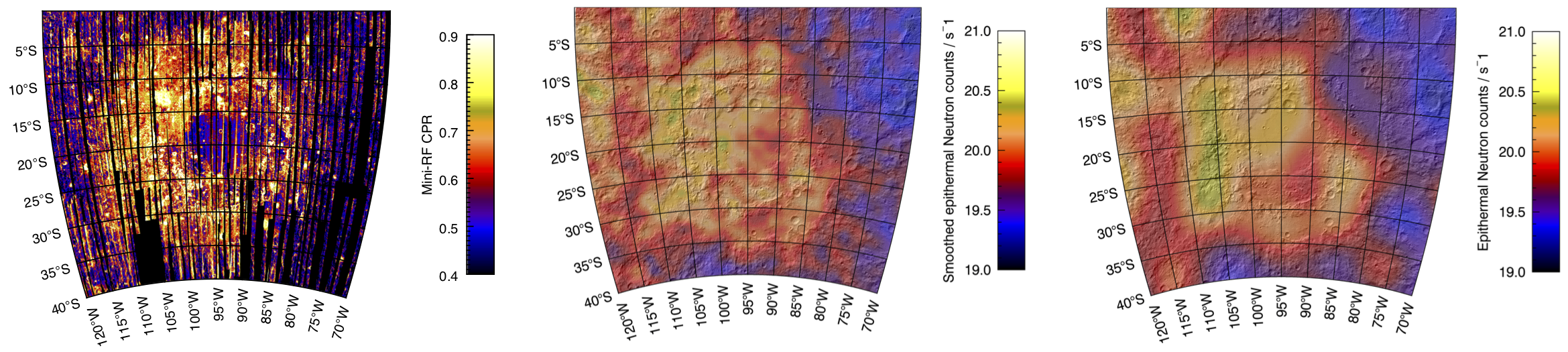}
\end{center}
\caption{{\it Left}: Orientale basin seen in  Mini-RF S-band circular polarization ratio. {\it Centre}: PSF-smoothed epithermal neutron reconstruction at Orientale basin. {\it Right}: Maximum Entropy Pixon epithermal neutron reconstruction at Orientale basin, superimposed on a LOLA shaded relied image.}
\label{fig:Orientale}
\end{figure*}

Another measure of surface maturity is Circular Polarization Ratio (CPR), defined as the ratio between power reflected in the same sense of circular polarization as that transmitted and the reflected echo in the opposite sense of circular polarization.  CPR is strongly enhanced by the presence of surface or near-subsurface roughness on the scale of the transmitted signal's wavelength.  Typically, as near-subsurface rockiness decreases with age so does CPR. However, Orientale has an age of around 3.8 Ga \citep{Neukum2001} and as expected shows OMAT and albedo consistent with this Late Imbrian age \citep{Cahill2014}.  However, LRO Miniature Radio Frequency (Mini-RF) S-band (12.6 cm) radar CPR is consistently immature in most parts of the basin.  The western part of the basin between Montes Rook and Cordillera exhibit some of the highest regional CPR values ($>$ 0.7) on the lunar surface (Figure~\ref{fig:Orientale}). \citet{Cahill2014} concluded on the basis of both these data sets and the lack of a substantial Diviner Rock Abundance (RA) signature in Orientale that the likely cause of varying degrees of contradiction between different maturity measures was the result of higher levels of roughness in the near subsurface on the scale of the S-band radar wavelength.

The original smoothed epithermal neutron data \citep{Maurice2004,Lawrence2015} are shown in Figure~\ref{fig:Orientale} (center) and the reconstructed data are shown in Figure~\ref{fig:Orientale} (right).  These maps illustrate the point described in section~\ref{ssec:mockResults} that for datasets with low SNR the pixon reconstruction minimizes spurious artifacts but preserves statistically robust features in the data.  The epithermal neutron data, which have a relatively low dynamic range compared with the thermal neutron and Th datasets, have many small-scale features in the smoothed map.  These small-scale features are all absent from the reconstructed map, which nonetheless contains all the large-scale features seen in the smoothed data.

When comparing the reconstructed epithermal neutron data to the noisy data it is interesting to note that at Orientale the maturity measures that sample the subsurface (CPR and epithermal neutrons) seem to conflict with those that are sensitive to the topmost-surface (normal albedo, OMAT and RA \citep{Cahill2015}).  \citet{Lucey2000} noted that the various types of maturity indices (e.g. albedo, soil grain size, OMAT) are determined by different factors (e.g. composition, density, depth, solar exposure) that are not necessarily correlated. Thus, the correlation of these indices, at a particular location, implies that the effective maturity is consistent across different time and depth scales. Conversely, as at Orientale, conflicting indices imply a more complex history or exposure layering scenario.  The epithermal neutron signature may be a consequence of poor solar wind implantation in the  blockier material implied by CPR to be present in the subsurface in western Orientale, due to the material's lower surface-to-volume ratio leading to fewer potential bonding sites. Further exploration of this process requires a detailed numerical model that is beyond the scope of this paper.


\subsubsection{Feldspathic Highlands Terrane}

\begin{figure*}
\begin{center}
\includegraphics[width=1.0\textwidth]{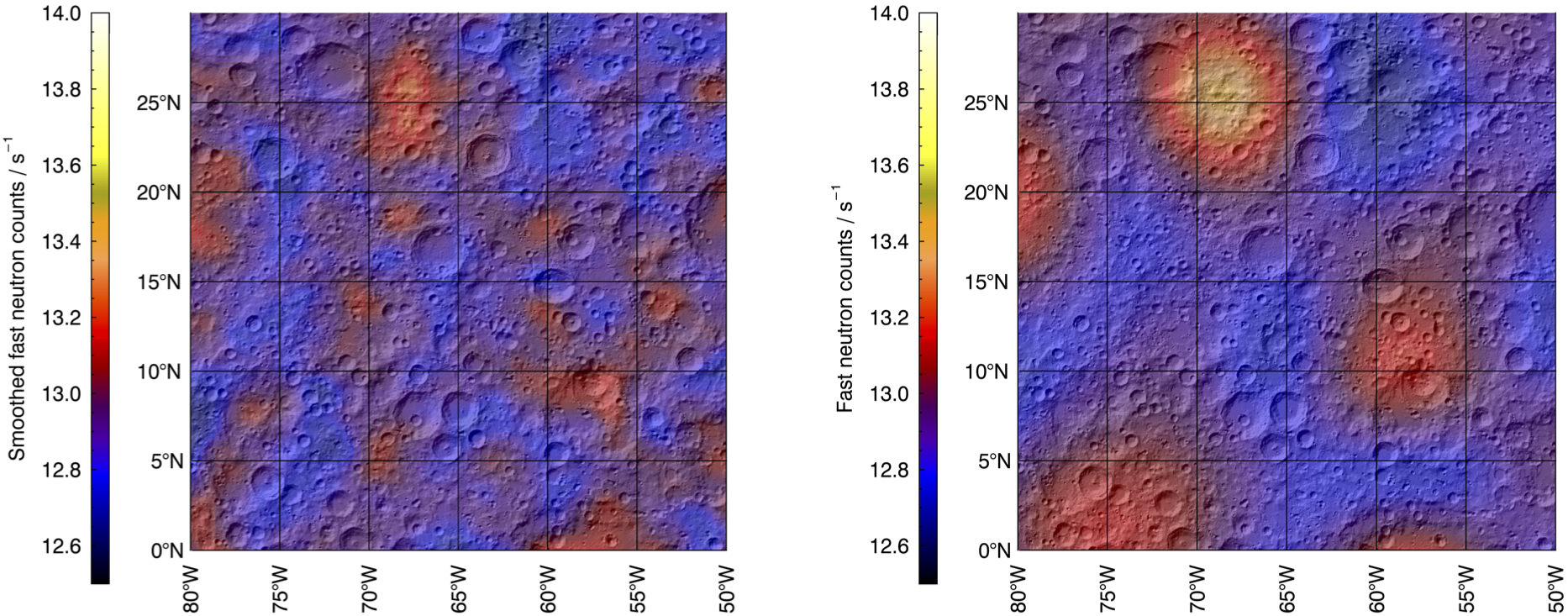}
\end{center}
\caption{A region of the Feldspathic Highlands Terrane with approximately uniformly low Fe content seen in ({\it left}) PSF-smoothed LPNS fast neutron data and ({\it right}) a Maximum Entropy Pixon reconstruction of the LPNS fast neutron data.}
\label{fig:Highlands}
\end{figure*}

The iron content of the central feldspathic highlands terrane (FHT) is nearly uniformly low \citep{Jolliff2000}.  Fast neutron flux is a measure of the mean atomic mass of the lunar surface as heavier nuclei produce more neutrons through nuclear spallation reactions.  As such, on the Moon fast neutrons trace the abundances of Fe and Ti \citep{Maurice2000,Gasnault2001}, with higher fast neutron flux corresponding to higher Fe concentrations.  

We have chosen this portion of the FHT to illustrate the effects of image reconstruction on a region of the surface with little signal contrast, to show how the noise in the data is reduced.  The unreconstructed, PSF-smoothed fast neutron data and pixon reconstruction of these data are shown in Figure~\ref{fig:Highlands}.  These smoothed data show variations with a small dynamic range on the spatial scale of the PSF,  in the reconstruction only broad variation is seen. As was also seen with the epithermal neutron data, this map illustrates how even in the absence of an increase in the spatial resolution of the data Bayesian image reconstruction improves the utility of the data by removing artifacts caused by the presence of noise.

\subsubsection{Th abundances at Copernicus crater}
Copernicus is a 93 km diameter rayed crater in eastern Oceanus Procellarum.  Dating of Copernicus ejecta returned by Apollo 12 revealed that Copernicus originated 800 Ma ago, consistent with its extensive ray system and pristine shape.  The crater has been noted as containing unusually low concentrations of Th based on LPGRS data \citep{Lawrence2003} and low FeO concentrations in Clementine reflectance observations.  Both of these observations are explained by the excavation of anorthositic, lower Th material.  

The reconstructed Th map shows significantly more structure than the unreconstructed map from \citet{Lawrence2003}.  We note (but do not show) that this reconstructed map also shows more structure than a prior Pixon reconstructed map of \citet{Lawrence2007}.  The fact that this map shows more structure is likely due to the fact that the LAP method was used here, which allows for more freedom than the prior version of the algorithm.

\begin{figure*}
\begin{center}
\includegraphics[width=1.0\textwidth]{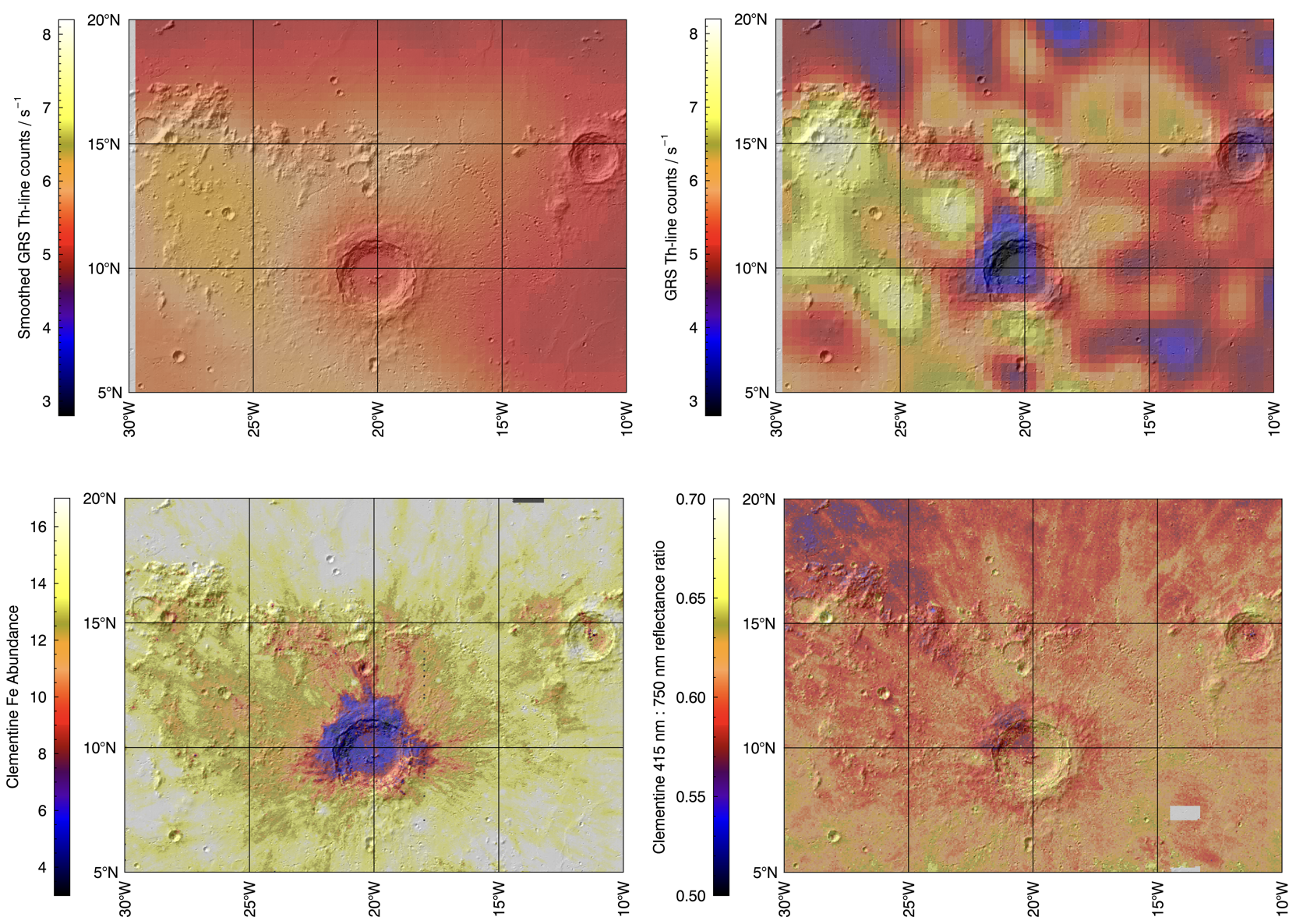}
\end{center}
\caption{{\it Top left}: PSF-smoothed Lunar Prospector GRS Th-line data at Copernicus crater. {\it Top right}: A Locally Adaptive Pixon reconstruction of the Lunar Prospector GRS Th-line data at Copernicus crater. Note that the color scale for the two Th maps is fixed at the same minimum and maximum values to illustrate the increase in dynamic range provided by the reconstruction. {\it Bottom left}: A Clementine-derived FeO concentration map at Copernicus crater. {\it Bottom right}: A Clementine spectral reflectance ratio map of the region around Copernicus cratershowing 414 nm/750 nm ratio. The low  415 nm / 750 nm ratio regions typically correspond to old ($\sim$4.5 billion years) highlands-like gabbroic anorthosite rocks.}
\label{fig:Copernicus}
\end{figure*}

In the low-Th region coincident with the crater, the smoothed map shows a distribution that is nearly symmetric around the crater (Figure~\ref{fig:Copernicus}). In contrast, the reconstructed map shows a distribution that is shifted northwest and has a non-circular shape that is similar to the low-FeO concentrations around the crater.

Finally, the Th data show that the low-Th region in Copernicus crater is coincident with both a small area exhibiting low Clementine 415 nm / 750 nm reflectance ratio in the northwestern quadrant of the crater and the diffuse low-FeO ejecta, which stretches to the east.  However, the agreement seems to be better with the Clementine albedo ratio feature.  This albedo ratio feature has been suggested to demarcate a `red spot', i.e. a region with high albedo and strong ultraviolet absorption \citep{Hagerty2015}.  These `red spots' are often interpreted to be of evolved silicic composition. They are therefore expected to be associated with high Th concentrations, as Th ions are incompatible and preferentially remain in a melt as it evolves.  However, the coincidence of the low-Th and high 750 nm / 415 nm reflectance ratio region implies that the material excavated by the Copernicus impact was composed of typical lunar highlands minerals not highly evolved rhyolite/granite.

\section{Conclusions}\label{sec:conc}

We have investigated, using mock data sets, several image reconstruction and deconvolution techniques commonly used with planetary gamma-ray and neutron data.  Using mock datasets, we found that the pixon methods are most effective at improving the resolution of, and suppressing noise in, these data at all SNRs and for all PSF footprint sizes.  However, the resolution improvement achieved was found to be dependent on both the SNR and the structure of the image being reconstructed. Reconstruction of low-SNR, images with few distinct features yields only a reduction in noise.

We applied the pixon image reconstruction methods to the LP NS and GRS Th-line data.  The improved resolution, achieved via image reconstruction, allows for direct identification of features in the NS/GRS datasets and higher-resolution spectroscopic and imaging data.  We saw that the high Clementine 750 nm / 415 nm reflectance ratio feature in Copernicus matches the low-Th region, revealing a FHT-like composition in the centre of the Procellarum KREEP Terrane.  Similarly, improving the resolution allows the identification of high albedo regions within Hertzsprung and Schr{\"o}dinger basins with high thermal neutron flux.




%
\appendix

\section{Misfit statistics used in Pixon reconstruction}
\subsubsection{Global misfit statistic, ${\bf E_R}$}\label{ssec:E_R}
${\bf E_R}$ is defined using as 
\begin{equation}
E_R = \sum_{{\bf y} = 0}^mA_{R}({\bf y})^{2},
\end{equation}
where $A_R$ is the autocorrelation of the residuals, $A_R({\bf y})
= \int R^*({\bf x}) R({\bf x} + {\bf y}){\rm d}{\bf x}$ for a pixel separation, or lag, of ${\bf y}$. And the reduced residuals between the
data and the blurred model are,
\begin{equation}\label{eqn:res}
R({\bf x}) = \frac{D({\bf x}) - (\hat{I}*B)({\bf x})}{\sigma({\bf x})},
\end{equation}
where $\sigma({\bf x})$ is the anticipated statistical noise in pixel
${\bf x}$.

Minimizing $E_R$ instead of $\chi^2$ helps suppress spatial
correlations in the residuals, which prevents spurious features
being formed by the reconstruction process.

\subsubsection{Local misfit statistic, ${\bf E^\prime_R}$}\label{ssec:LocalE_R}
The local $E_R$ statistic, ${\bf E^\prime_R}$, was created via localizing $A_R$ by multiplying by a kernel, $K$, such that
\begin{equation}
A^\prime_R({\bf x},{\bf z}) = \int R({\bf y}) R({\bf x} + {\bf y}) K({\bf z} - {\bf y}) \, d{\bf y},
\end{equation}
which we can rewrite in a form resembling a convolution, by defining $C_{\bf x}({\bf y}) = R({\bf
y})R({\bf x} + {\bf y})$, as
\begin{equation}\label{eqn:A_R_prime}
A^\prime_R({\bf x},{\bf z}) = \int C_{\bf x}({\bf y}) K({\bf z} - {\bf y}) \, d{\bf y}.
\end{equation}
For computational simplicity we select a Gaussian kernel.  The kernel smoothing length is chosen to be larger than the largest pixon width so that it is the pixon size that sets the smoothing scale in the reconstruction.

\section*{Acknowledgements}
This work was supported by a grant from the the Lunar Advanced Science and Exploration Research Program (grant NNX13AJ61G) and used the DiRAC Data Centric system at Durham University,
operated by the Institute for Computational Cosmology on behalf of the
STFC DiRAC HPC Facility (www.dirac.ac.uk). This equipment was funded
by BIS National E-infrastructure capital grant ST/K00042X/1, STFC
capital grant ST/H008519/1, and STFC DiRAC Operations grant
ST/K003267/1 and Durham University. DiRAC is part of the UK national
E-Infrastructure.


\bibliographystyle{elsarticle-harv}
\bibliography{refs.bib}

\end{document}